\documentclass[aps,prl,twocolumn,showpacs,floatfix,superscriptaddress]{revtex4}
\usepackage[latin1]{inputenc}
\usepackage{bbm,slashed}
\usepackage{graphicx,epsfig}
\usepackage{amsmath,amsfonts,amssymb}  

\usepackage{hyperref}
\usepackage[active]{srcltx}

\usepackage{color}
\definecolor{holger}{rgb}{0,0.4,0.7}
\definecolor{comment}{rgb}{0.9,0,0}
\definecolor{alexej}{rgb}{0.5,0,0.5}

\definecolor{alexej2}{rgb}{0.0,0.5,0.5}

\newcommand{\pt}{\mathcal{T}}

\newcommand{\bgam}{\boldsymbol{\gamma}}

\newcommand{\Eqref}[1]{Eq.~\eqref{#1}}

\DeclareMathAlphabet{\boldmathe}{T1}{cmr}{bx}{it}

\begin{document}
\title{Non-monotonic thermal Casimir force from geometry-temperature interplay} 
\author{Alexej Weber}
\affiliation{
\begin{minipage}{16cm}
Institut f{\"u}r Theoretische Physik, Universit{\"a}t Heidelberg, 
Philosophenweg 16, D-69120 Heidelberg, Germany
\end{minipage}}
\author{Holger Gies}
\affiliation{
Theoretisch-Physikalisches Institut, Friedrich-Schiller-Universit{\"a}t
Jena, Max-Wien-Platz 1, D-07743 Jena, Germany}

\begin{abstract}
  The geometry dependence of Casimir forces is significantly more
  pronounced in the presence of thermal fluctuations due to a generic
  geometry-temperature interplay. We show that the thermal force for standard
  {sphere-plate or cylinder-plate geometries develops a}
  non-monotonic behavior already in the simple case of a fluctuating Dirichlet
  scalar. In particular, the attractive thermal force can {\em increase} for
  increasing distances below a critical temperature. This anomalous
  behavior is triggered by a reweighting of relevant fluctuations on the scale
  of the thermal wavelength.
The essence of the phenomenon becomes transparent within the
worldline picture of the Casimir effect.

\end{abstract}

\maketitle

Fluctuations of the radiation field between mesoscopic or macroscopic test
bodies give rise to the fascinating Casimir effect \cite{Casimir:dh} -- a
dispersive quantum and relativistic force phenomenon in the absence of net
charges, see \cite{Milton:2001yy} for reviews. Experimental verifications
\cite{Lamoreaux:1996wh} typically involve spheres or cylinders and plates. For
standard materials, the Casimir force is generally attractive
\cite{Kenneth:2006vr} and decreases monotonically with distance. The latter
seems intuitively clear from spectral properties of the fluctuations:
in this picture, the Casimir effect arises from the difference between the
fluctuation spectrum in the presence of the surfaces and that of the trivial
vacuum (at infinite surface separation). For increasing 
separation, the spectrum is expected to monotonically approach the vacuum
spectrum, implying a monotonic force depletion.

A first non-monotonic behavior has been observed in a more involved
piston-like geometry of two squares moving between metal walls
\cite{Rodriguez:2007}; {similar observations hold for two cylinders near
  a sidewall \cite{Rahi}.} Here, the non-monotonic behavior arises from a
competition between the TE and TM modes of the electromagnetic
fluctuations. {Its strength is governed by the dependence of the force on
  a lateral geometry parameter.} The example demonstrates that an unexpected
behavior of the Casimir force may occur in the presence of competing scales
(in this case: normal and lateral distances).

In this work, we show that a non-monotonic behavior already exists for
a single fluctuating scalar obeying Dirichlet boundary conditions on the
surfaces (similar to a TM mode in a cavity-like
configuration). This anomalous phenomenon requires a nonzero temperature and
occurs for the thermal contribution to the Casimir force which we abbreviate
by ``thermal force'' in the following.

This phenomenon is a prime example of the general geometry-temperature or {\em
  geothermal} interplay \cite{Gies:2008zz}. As first conjectured in
\cite{Scardicchio:2005di}, the thermal correction to the Casimir effect can
vary qualitatively for different geometries, as both the zero- and
finite-temperature Casimir effect are based on the underlying fluctuation
spectrum. Analytical and numerical evidence for this geothermal interplay has
been collected using perpendicular- or inclined-plates
configurations \cite{Gies:2008zz,Weber:2009dp,Gies:2009nn}. At 
low temperatures, the temperature dependence is more pronounced in {\em open}
geometries which do not exhibit a gap in the relevant part of the fluctuation
spectrum. At any small value of the temperature, thermal modes can be excited,
inducing a characteristic thermal force. By contrast, geometries with a gap in
the spectrum show a suppressed thermal force, as thermal modes can hardly be
excited for temperatures below the gap scale.

Typical configurations used in experiments involve spheres or cylinders above
a plate which are open geometries without a spectral gap. Investigating
the geothermal interplay for these geometries therefore is an urgent
problem. For the fluctuating electromagnetic field, first results
for the thermal Casimir force in the sphere-plate configuration have recently
been obtained \cite{Canaguier:2009,Bordag:2009} using scattering
techniques. In the limit of temperature being smaller than both the inverse
sphere radius, $T\ll 1/R$, and the inverse sphere-plate distance, $T\ll 1/a$,
the thermal force in \cite{Bordag:2009} is always attractive for any value of
$a/R$ and monotonically decreasing with increasing separation $a$. By contrast,
the thermal force derived in \cite{Canaguier:2009} using a truncated multipole
summation shows a repulsive behavior for smaller distances and becomes
attractive at larger distances. From their data, a non-monotonic behavior of
the thermal force at larger distances $a/R\gg 1$ and low temperatures $T
R\lesssim 1$ can be anticipated. Whereas both studies nicely agree in the
limit of low temperature and small spheres, the seeming disagreement beyond
the strictly asymptotic validity regimes of the two different expansions requires clarification.

In this work, we demonstrate that non-monotonic thermal forces indeed occur
unambiguously for the sphere-plate and cylinder-plate configuration in the
low-temperature region. For simplicity and clarity, we use Dirichlet scalar
fluctuations in order to avoid additional complications from competing
{polarization} modes. Most importantly, the occurrence of this anomalous
behavior can be understood as a temperature-induced reweighting of different relevant
fluctuations in a given geometry. This is at the heart of the geothermal interplay.

\begin{figure}[t]
\begin{center}
\includegraphics[width=0.8\linewidth]{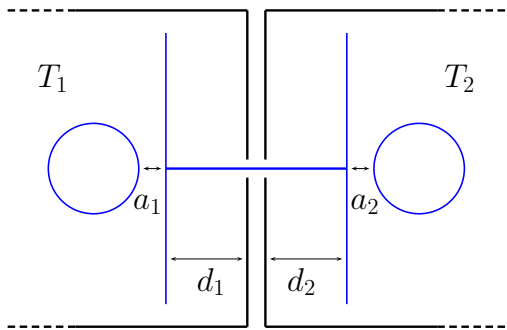}
\end{center}
\caption{Setup for a measurement of the thermal force for the sphere-plane
  configuration: by tuning the sphere positions to identical distances
  $a_1=a_2$, the zero-temperature force on the tightly connected pair of plates
  cancels. Placing the two sphere-plate subsystems into two heat baths at
  different temperatures $T_1$ and $T_2$, the different thermal forces induce
  a net force on the pair of plates. We assume that all distances between the
  (blue) Casimir surfaces and the (black) heat-bath boundaries are large, e.g.,
  $d_{1,2}\gg a_{1,2}$, such that corresponding Casimir interactions can be
  neglected.  
}\label{F1}
\end{figure}

The origin of this phenomenon can most intuitively be understood in the
worldline picture of the Casimir effect \cite{Gies:2003cv}. The worldline
formalism 
{maps} the standard spectral Casimir
problem to a 
{Feynman} path integral over closed worldlines in
position space. These worldlines can be viewed as the space trajectories of
quantum fluctuations within a given Casimir geometry. In addition to 
this intuitive picture, the worldline approach offers a powerful
tool for numerical as well as analytical studies. 
E.g., for the sphere/cylinder-plate configuration at zero temperature, a coherent
understanding of the Casimir force has emerged from worldline studies
\cite{Gies:2006bt} on the one hand and scattering techniques on the other hand
\cite{Bulgac:2005ku,Emig:2006uh,Bordag:2006vc,Milton:2007gy}. A similar
consensus has been reached for ``edge effects''
\cite{Gies:2006xe,Weber:2009dp}. Also for more involved geometries, the
worldline picture directly provides for a qualitative understanding \cite{Schaden:2009zz}.

The Casimir free energy induced by a fluctuating Dirichlet scalar reads for
any configuration \cite{Gies:2003cv,Gies:2006bt}
\begin{eqnarray}\label{Int-3}
E_\mathrm{c}&=&-\frac{1}{32 \pi^{2}}  
\int_{0}^\infty \frac{\mathrm{d}\pt}{\pt^{3}}\!\!\sum_{n=-\infty}^{\infty}
\!\!e^{-\frac{n^2\beta^2}{4 \pt}}\!\! \int_{\mathbf{x}_{\text{CM}}}  
\left\langle  \Theta [\bgam;\pt]
\right\rangle.
\end{eqnarray}
Here, $\langle \dots \rangle$ denotes an average over an ensemble of
worldlines $\bgam$  with Gau\ss ian velocity distribution with a common
center of mass $\mathbf{x}_{\text{CM}}$. The auxiliary propertime parameter $\pt$ acts as a spatial scaling
factor and governs the size of the worldlines. The sum over winding number $n$
counting the round trips of a worldline around the finite-temperature torus
takes care of thermal fluctuations, with the $n=0$ term corresponding to the
zero-temperature $T=1/\beta\to0$ result. The $\Theta$ functional obeys
$\Theta=1$ if a given worldline intersects two or more interacting surfaces, and
$\Theta=0$ otherwise.

\begin{figure}[t]
\begin{center}
\includegraphics[width=0.8\linewidth]{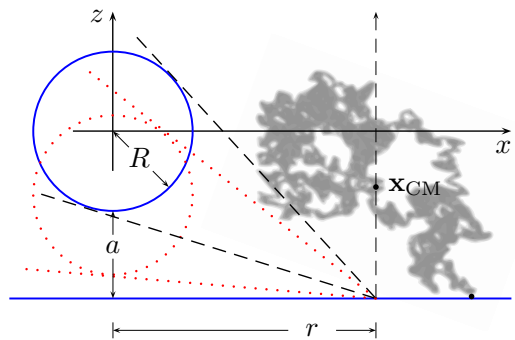}
\end{center}
\caption{Sketch of the sphere-plate configuration. 
  During the
  propertime integration, the worldline is always attached to the plate, while
  all its points move on rays originating from the projection of its center of
  mass $\mathbf{x}_{\text{CM}}$ on the plate at polar coordinate $r$.  Only
  points lying inside a cone wrapping around the sphere with the tip of the
  cone at $r$ (thick dashed lines) pass through the sphere for increasing
  $\pt$ and thus contribute to the
  Casimir force.  In general, for every $r\gg R$ there exists a maximal
  separation $a$, such that the sphere can be intersected by the worldline. If
  the sphere separation is reduced (dotted red circle), it can become
  invisible for a given worldline, such that the worldline stops
  contributing to the Casimir force. This induces the phenomenon that
  the thermal force can increase for increasing separation. }\label{F2}
\end{figure}

In the following, we are exclusively interested in the thermal contribution
$n\neq 0$, serving as the interaction potential for the thermal force, $F_T=-d
E_{\mathrm{c}}/da|_{n\neq 0}$; here, $a$ is a distance parameter between
disjoint bodies. For instance, for a sphere above a plate, a configuration
that allows to cancel the zero-temperature forces is sketched in
Fig.~\ref{F1}. Alternatively, the zero-temperature force could be balanced by
applying suitable electrostatic potentials to a single-sphere setup. We
  stress that the  thermal-force phenomena discussed in the following are
  dominated by the corresponding zero-temperature forces in the standard setups where
    $a\ll R$. After removing the zero-$T$ contribution, the thermal force
reads
\begin{align}\label{CEff-SaP-4}
F_T=-\frac{1}{{8}\pi} \sum_{n=1}^\infty  \int_{0}^\infty \!\!
\mathrm{d}\mathcal{T} \frac{e^{-\frac{n^2 \beta^2}{4\pt}}}{\mathcal{T}^{3}}
\int_0^\infty \mathrm{d}r \, r\, \left\langle \Theta_{\mathrm{S}} \right\rangle,
\end{align}
where $r$ is the polar center-of-mass coordinate on the plate. 
$\Theta_{\mathrm{S}}$  measures whether a worldline which is attached to
the plate also intersects with the sphere for certain scaling sizes $\sim
\pt$, see Fig.~\ref{F2}. For the cylinder-plate case, the factor of $r$ has to
be replaced by the (infinite) length $L_y$ of the cylinder {divided by $\pi$}.

The geothermal effect can easily be understood in the worldline picture: at
zero temperature, the Casimir force is generically dominated by small
propertimes, i.e., small worldlines with a minimal extent such that the
worldlines can intersect with both surfaces. As a consequence, the energy
density is typically peaked in the region near minimal separation. By
contrast, the peak of the finite-temperature propertime factor $\sim e^{{-}n^2
  \beta^2/(4\pt)}/\pt^3$ moves to larger $\pt$ values for decreasing
temperature. Therefore, as larger worldlines can contribute, the free-energy
density is potentially distributed over a larger region of space. Whether or
not this broadening occurs depends on the details of the geometry, as
worldlines at larger distances still have to intersect the Casimir surfaces. 

\begin{figure}[t]
\begin{center}
\includegraphics[width=0.9
\linewidth]{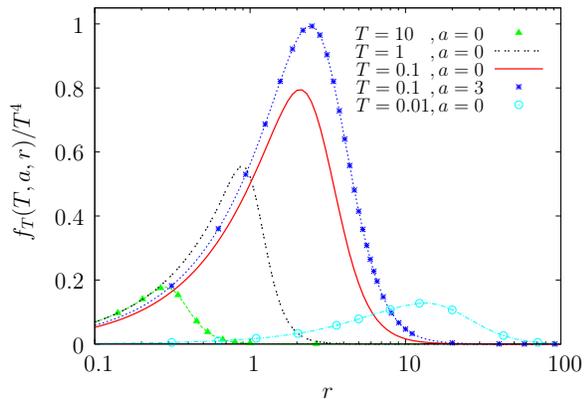}
\end{center}

\vspace{-0.4cm} 

\caption{{Radial} thermal force density $f_T(T,a,r)$ for a sphere above a
  plate for different temperatures $T$ {in units of $R=1$. The peak position as
  a function of $r$ increases with decreasing temperature.} The statistical
  error is below one percent.  }\label{F3}
\end{figure}

In Fig. \ref{F3}, we plot the radial distribution of the thermal 
force density for various temperature values. 
Its peak position
increases with decreasing temperature.  
This corresponds to the fact that low temperatures can still excite
long-wavelength modes if the spectrum is not gapped. Figure \ref{F3} also
demonstrates that any local approximation of the Casimir force such as the
proximity force approximation (PFA) is generically bound to fail for a proper
description of the geometry-temperature interplay; see, however, 
\cite{Milton:2009gk} for semitransparent surfaces. For experiments at low
temperature, our results indicate that the idealized sphere-plate
configuration requires the plate to be much larger than the sphere. Otherwise
thermal edge effects have to be accounted for, being more severe than edge
effects at $T=0$.

In addition to the potential delocalization of the thermal force
density, a second geometric mechanism is operative for the non-monotonic
behavior of the thermal Casimir force: in the worldline picture formalized by
\Eqref{CEff-SaP-4}, the Casimir force arises from all worldlines that are
attached to the plate and intersect the sphere. Consider a worldline with
center of mass polar coordinate $r$ in the sphere-plate geometry at separation
$a$, see Fig.~\ref{F2}. Upon integrating over propertime $\pt$, the size of the worldlines are
scaled but the center of mass projection onto the sphere stays fixed at
$r$. Therefore, the worldline points that eventually intersect the sphere have
to lie inside a cone with its tip attached to the plate at $r$, wrapping
around the sphere, see Fig.~\ref{F2}. 

Let us now reduce the sphere-plate separation $a$ for a given worldline. As
this moves the corresponding cone towards the plate (red dotted lines in
Fig.~\ref{F2}), all points of a given worldline may drop out of the cone such
that this worldline no longer contributes to the Casimir force. This is the
mechanism that potentially {\em reduces} the Casimir force for smaller
separations. 

While this geometric argument is independent of any temperature, this loss
mechanism of relevant fluctuations is negligible at zero temperature:  the
effect is outweighed by small worldlines intersecting sphere and plate near
the point of closest separation which dominate the zero-temperature Casimir
force. By contrast, the finite-temperature propertime factor $\sim e^{{-}n^2
  \beta^2/(4\pt)}/\pt^3$ favors the contribution of larger worldlines at low
temperature and thus emphasizes the relevance of the loss mechanism at larger
separations. Whether or not a non-monotonic thermal force law arises then is
a competition between small worldlines in the region of close separation
  and large worldlines on the scale of the thermal wavelength.
If the contribution of the latter to the thermal force is
dominant, the thermal force can increase for increasing distance as more and
more worldlines can contribute, i.e., become relevant fluctuations.

\begin{figure}[t]
\begin{center}
\includegraphics[width=0.9
\linewidth]{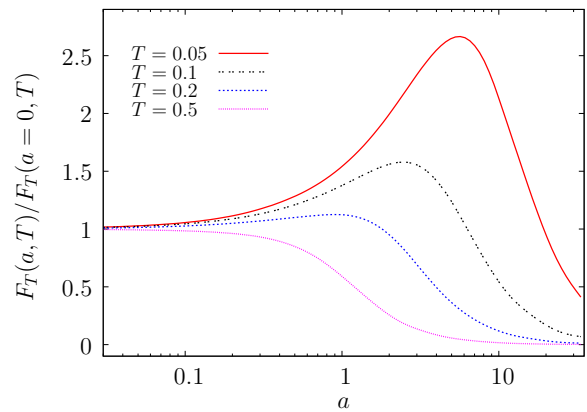} 
\end{center}

\vspace{-0.4cm} 

\caption{Thermal Casimir force of a sphere for various
  temperatures $T$ and $R=1$, normalized to the thermal force at $a=0$. For
  sufficiently small temperatures, the absolute value of the thermal force
  {$F_T$} first \textit{increases} with increasing $a$. For $T\lesssim
  0.05$ the small-$a$ behavior of the normalized curves are well described by
  $1+a(2.68 R T^4- R^2 15.7\,T^5)/F_T(a=0,T)$. 
  The statistical error is on the order of the curve thickness.
}\label{FT-SaP-F4}
\end{figure}

Worldline numerical results for the thermal force for a sphere above a plate
as a function of separation $a$ are displayed in
Fig.~\ref{FT-SaP-F4}. {We have used up to $40\, 000$ worldlines discretized by $2\cdot 10^6$ points each.} As the
limit of zero separation $a\to0$ exists for thermal forces according to a
general argument \cite{Gies:2009nn}, we normalize the thermal force relative to the
$a\to0$ limit. All dimensionful scales are normalized to the sphere radius
which we set to $R=1$ in the following. For temperatures larger than a
critical temperature $T>T_{\text{cr}}\simeq {0.34(1)/R}$, we observe an attractive
thermal force which decreases monotonically for increasing sphere-plate
separation $a$ and thus satisfies standard expectations. For smaller
temperatures $T<T_{\text{cr}}$, the thermal force first increases for
increasing separation, develops a maximum and then decreases to zero for
infinite separation. The peak position increases with increasing thermal
wavelength, i.e., inverse temperature. In all cases, the force remains
attractive. {As an example, room temperature $T=300$K is critical for
  spheres of radius $R\simeq 2.6\mu$m, supercritical for larger and
  subcritical for smaller spheres. At $T=70$K and $R=1.6\mu$m, the thermal
  force increases up to $a\simeq 9\mu$m. }

For small $T\ll 1/R$  and $a\ll R$, our numerical data is compatible with an
expansion of the type
\begin{equation}\label{FT-SaP-1}
{F_T}= c_2 R^2 T^4 +  c_3 a R T^4 +
\mathcal{O}\left({(a/R)^2,(TR)^5}\right), 
\end{equation}
where $c_2\approx-3.96(5)$ and $
c_3\approx -2.7(2)$. The absence of terms $ \sim c_0 R T^3$ or $c_1 aT^3$ is
in agreement with scaling arguments \cite{ToAppear} and numerically confirmed
in the regime $T>0.01$ where data is available.

Since $c_2$ and $c_3$ have the same sign, the absolute value of the thermal
Casimir force $F_T$ \textit{increases} with increasing $a$ for sufficiently
small $a$ and $T$.  In passing, we note that the high-temperature limit $T\gg
1/R$ agrees with the PFA prediction for $a\to 0$: $F_T=-\zeta(3)R T^3/2$.  At
low temperatures, the PFA disagrees with the full result even in the $a\to0$
limit, as observed in \cite{Gies:2009nn}.

The case of a cylinder above a plate is different from the sphere-plate case
in two respects: first, the missing polar measure factor $r$ reduces the
weight of distant worldlines. Second, the probability for a given worldline to
intersect the infinitely long cylinder is larger than for a sphere.  We
observe that these two effects balance each other, leading again to a $T^4$
behavior at low $T$: 
\begin{align}\label{FT-CaP-1}
{F_T}
/{L_y}= c_2 R  T^4 +  c_3 a T^4 + \dots,
\end{align}
where $c_2\approx -1.007(7)$ and $c_3\approx -0.41(4)$. The coefficients again have the same sign, {implying} that the thermal
force increases with $a$ for sufficiently small $a$ and $T$. The behavior is
then similar to Fig.~\ref{FT-SaP-F4}. For the critical temperature, we obtain
$T_\mathrm{cr}\approx 0.31(1)/R$.  At high temperatures and $a=0$, the thermal
force agrees with the PFA prediction as expected, ${F_T}\approx -0.278
\sqrt{R} T^{\frac{7}{2}}$.

To summarize, we have shown that the thermal contribution to the Casimir
  force for the standard sphere-plate and cylinder-plate geometry develops a
  non-monotonic behavior already in the simple case of a fluctuating
  Dirichlet scalar. For temperatures below a critical value and for small
  distances, the thermal force increases for increasing separation. This
  phenomenon is a prime example for the geometry-temperature interplay in open
  geometries. It has a simple geometric interpretation in the worldline
  picture and can reliably be computed in the numerical worldline approach. 

  We stress that the case of electromagnetic fluctuations can develop yet
  another thermal-force behavior due to possibly competing polarization modes.
  For a quantitative comparison with experimental data, a nontrivial interplay between
  geometry, temperature as well as dielectric material properties can be
  expected; pioneering first results on this interplay have been obtained in
  \cite{Canaguier:2009}.

\acknowledgments{We thank T.~Emig {for suggesting the setup in Fig.~\ref{F1}}  and E.~Elizalde for useful discussions. We
  have benefited from activities within ESF Research network CASIMIR. We
  acknowledge support by the Landesgraduiertenf\"orderung
  Baden-W\"urttemberg, by the Heidelberg Graduate School of Fundamental
  Physics, and by the DFG under Gi 328/3-2 and  Gi 328/5-1.  }

\end{document}